\pdfoutput=1

\documentclass[onecolumn,12pt,preprintnumbers,amsmath,amssymb,floatfix]{revtex4}

\usepackage{graphicx}
\usepackage{dcolumn}
\usepackage{bm}
\usepackage{amsfonts}

\DeclareMathAlphabet{\mathsfsl}{OT1}{cmr}{bx}{it}
\begin{document}
\title{Cooling under applied stress rejuvenates amorphous alloys and enhances their ductility}
\author{Nikolai V. Priezjev$^{1,2}$}
\affiliation{$^{1}$Department of Mechanical and Materials
Engineering, Wright State University, Dayton, OH 45435}
\affiliation{$^{2}$National Research University Higher School of
Economics, Moscow 101000, Russia}
\date{\today}
\begin{abstract}

The effect of tensile stress applied during cooling of binary
glasses on the potential energy states and mechanical properties is
investigated using molecular dynamics simulations. We study the
three-dimensional binary mixture that was first annealed near the
glass transition temperature and then rapidly cooled under tension
into the glass phase. It is found that at larger values of the
applied stress, the liquid glass former freezes under higher strain
and its potential energy is enhanced.  For a fixed cooling rate, the
maximum tensile stress that can be applied during cooling is reduced
upon increasing initial temperature above the glass transition
point. We also show that the amorphous structure of rejuvenated
glasses is characterized by an increase in the number of contacts
between smaller type atoms. Furthermore, the results of tensile
tests demonstrate that the elastic modulus and the peak value of the
stress overshoot are reduced in glasses prepared at larger applied
stresses and higher initial temperatures, thus indicating enhanced
ductility. These findings might be useful for the development of
processing and fabrication methods to improve plasticity of bulk
metallic glasses.

\vskip 0.5in


\end{abstract}

\maketitle

\section{Introduction}

Understanding the relation between amorphous structure and
mechanical and physical properties of disordered solids is important
for numerous biomedical and structural
applications~\cite{Khan18,ZhengBio16}.  Bulk metallic glasses are
known to posses advantageous properties, including high yield stress
and large elastic limit, but they become brittle when in a relaxed
state~\cite{Schuh03,Ma12,ShangWang14,Priez20tfic,Priez20star,Ozawa20}.
To increase ductility, metallic glasses can be rejuvenated via
thermal treatment or mechanical agitation~\cite{Greer16}.  In
general, the processing methods that are used to control the energy
states of amorphous alloys can be broadly distinguished in two
categories based on their temperature relative to $T_g$. The first
one deals with either cooling from the liquid state into the glass
phase or flash annealing where a glass is temporarily brought to the
liquid state and then cooled below $T_g$ with a suitably fast
rate~\cite{Ogata15,Maass18,Priez19one,WangTang20}. Other methods
involve mechanical and thermal manipulation of amorphous materials
within the glass phase. Common examples include ion
irradiation~\cite{Gianola14,Xue19}, high-pressure
torsion~\cite{Meng12,Langdon12}, shot peening~\cite{Mear08,Sort13},
elastostatic loading~\cite{Park08,Jae-Chul09,Tong13,Wang15,Bai15,
GreerSun16,Zhang17,PanGreer18,PriezELAST19, PriezELAST21}, and
thermal cycling~\cite{Ketov15,Guo19,Priez18tcyc,
Priez19T2000,Priez19T5000,Mirdamadi19,Jittisa20,Meylan20,Du20} among
others. On the other hand, amorphous alloys can be mechanically
annealed to lower energy states by applying time periodic
deformation with an amplitude below the yielding
point~\cite{Lacks04,Sastry13,Priezjev18,NVP18strload,Jana20}, and
the decay of the potential energy is accelerated when the loading
orientation is periodically alternated in two or three spatial
dimensions~\cite{NVP19alt,NVP20altY}.  Most recently, it was
discovered experimentally that upon cooling under tension, bulk
metallic glass formers become highly rejuvenated and their bending
ductility can be tripled~\cite{Schroers20}. These findings were
rationalized by invoking the concept of fictive temperature upon
cooling when the characteristic timescale of structural relaxation
is comparable to the inverse cooling rate. Moreover, the fictive
temperature can be increased by the application of sufficiently high
strain rate during freezing, which leads to enhanced
ductility~\cite{Schroers20}. However, despite recent progress,
optimization and design of the cooling protocols, including stress-
versus strain-controlled deformation as well as the range of
annealing temperatures and applied stresses, remain a challenging
problem.

\vskip 0.05in

In this paper, the influence of initial temperature and tensile
stress applied during the cooling process on rejuvenation and
mechanical properties of binary glasses is studied via molecular
dynamics simulations. We consider a binary mixture initially
annealed at temperatures slightly above the glass transition
temperature and then rapidly cooled under applied tensile stress
deep into the glass phase. It will be shown that with increasing
tensile stress, glasses are relocated to higher energy states and
become more ductile. Moreover, for a given cooling rate, the initial
temperature determines the maximum value of the applied stresses and
therefore the maximum strain rate during freezing.

\vskip 0.05in

The rest of the paper proceeds as follows. The parameters of the
interaction potential and the preparation procedure as well as the
protocol for cooling under stress are described in the next section.
The results for the system deformation, potential energy, atomic
structure, and mechanical properties are presented in
section\,\ref{sec:Results}. The main conclusions are briefly
summarized in the last section.

\section{Molecular dynamics simulations}
\label{sec:MD_Model}

In our study, the model glass is represented via a mixture of two
types of atoms with strongly non-additive interaction that prevents
crystallization when the system is cooled below the glass transition
temperature~\cite{KobAnd95}. This model was first introduced by Kob
and Andersen (KA) about twenty years ago and has since been
extensively studied in the context of glass transition, dynamical
heterogeneity, and yielding in amorphous materials~\cite{KobAnd95}.
It should be noted that parameters of the KA binary mixture model
are similar to those used by Weber and Stillinger to study the
amorphous metal alloy $\text{Ni}_{80}\text{P}_{20}$~\cite{Weber85}.
In the KA model, the interaction between atoms types
$\alpha,\beta=A,B$ is described by the Lennard-Jones (LJ) potential
as follows:
\begin{equation}
V_{\alpha\beta}(r)=4\,\varepsilon_{\alpha\beta}\,\Big[\Big(\frac{\sigma_{\alpha\beta}}{r}\Big)^{12}\!-
\Big(\frac{\sigma_{\alpha\beta}}{r}\Big)^{6}\,\Big],
\label{Eq:LJ_KA}
\end{equation}
where $\varepsilon_{AA}=1.0$, $\varepsilon_{AB}=1.5$,
$\varepsilon_{BB}=0.5$, $\sigma_{AA}=1.0$, $\sigma_{AB}=0.8$,
$\sigma_{BB}=0.88$, and $m_{A}=m_{B}$~\cite{KobAnd95}. The total
number of atoms is $N=60\,000$. To alleviate the computational
burden, the cutoff radius of the LJ potential was fixed to
$r_{c,\,\alpha\beta}=2.5\,\sigma_{\alpha\beta}$.  The results are
reported in terms of the LJ units of length, mass, energy, and time,
as follows: $\sigma=\sigma_{AA}$, $m=m_{A}$,
$\varepsilon=\varepsilon_{AA}$, and, correspondingly,
$\tau=\sigma\sqrt{m/\varepsilon}$.  The MD simulations were
performed in parallel using the efficient LAMMPS code~\cite{Lammps}
with the integration time step $\triangle t_{MD}=0.005\,\tau$,
periodic boundary conditions, and the Nos\'{e}-Hoover
thermostat~\cite{Allen87}.

\vskip 0.05in


The first step in the preparation procedure was to equilibrate the
binary mixture near the glass transition temperature at zero
pressure. It was originally demonstrated, by extrapolating diffusion
coefficients of both types of atoms as a function of temperature,
that the critical temperature of the KA model at the constant
density $\rho=\rho_A+\rho_B=1.2\,\sigma^{-3}$ is
$T_c=0.435\,\varepsilon/k_B$~\cite{KobAnd95}. More recently, it was
found from the crossover of the potential energy curve, that the
glass transition temperature is about
$T_g\approx0.35\,\varepsilon/k_B$ when the system is cooled from the
liquid state with the rate $10^{-5}\varepsilon/k_{B}\tau$ at zero
pressure~\cite{Priez19one}. In the present study, the binary mixture
was first equilibrated at $T_{LJ}=1.0\,\varepsilon/k_B$ and then
annealed at four temperatures near $T_g$, namely,
$T_{LJ}=0.36\,\varepsilon/k_B$, $0.38\,\varepsilon/k_B$,
$0.40\,\varepsilon/k_B$, and $0.42\,\varepsilon/k_B$ during
$2\times10^5\,\tau$ in a periodic box at $P=0$. It was previously
shown that the binary mixture undergoes structural relaxation during
the time interval $2\times10^5\,\tau$ at
$T_{LJ}\geqslant0.32\,\varepsilon/k_B$, and the potential energy of
inherent structures decreases upon reducing annealing temperature
towards $T_g$~\cite{Priez20tfic}.

%
\begin{figure}[t]
\includegraphics[width=9.0cm,angle=0]{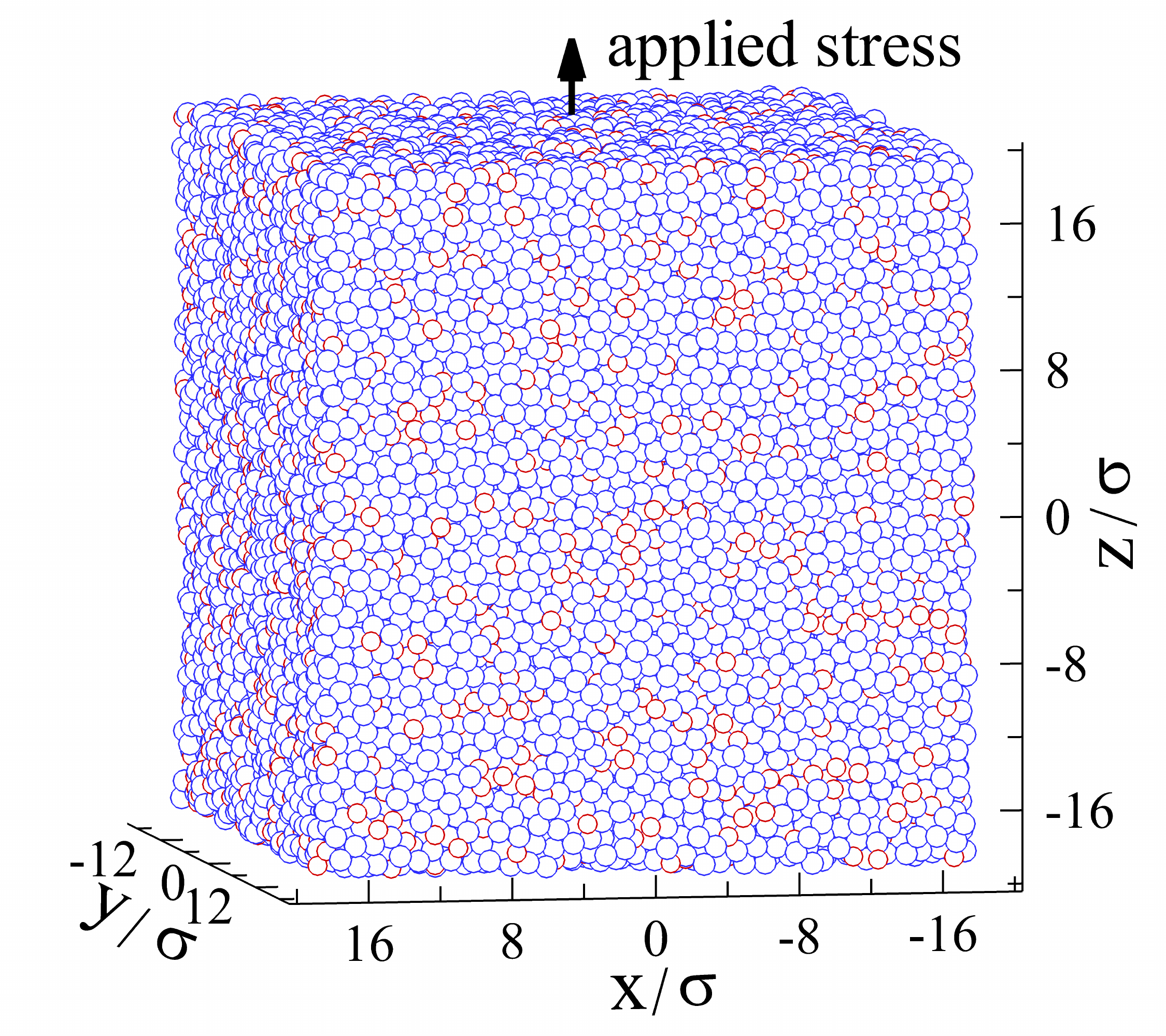}
\caption{(Color online) The snapshot of the binary glass after
cooling from $T_{LJ}=0.36\,\varepsilon/k_B$ to
$0.01\,\varepsilon/k_B$ with the rate $10^{-4}\varepsilon/k_{B}\tau$
and at the same time applying the constant stress
$0.6\,\varepsilon\sigma^{-3}$ along the $\hat{z}$ direction. The
stress components along the $\hat{x}$ and $\hat{y}$ directions are
set to zero. The total number of atoms is $60\,000$. The atoms of
types $A$ and $B$ are denoted by large blue and small red spheres.
The atoms are not drawn to scale.}
\label{fig:snapshot_system}
\end{figure}

\vskip 0.05in


Following the equilibration period, the system was linearly cooled
from the initial temperatures ($0.36\,\varepsilon/k_B$,
$0.38\,\varepsilon/k_B$, $0.40\,\varepsilon/k_B$, and
$0.42\,\varepsilon/k_B$) to $T_{LJ}=0.01\,\varepsilon/k_B$ with the
rate of $10^{-4}\varepsilon/k_{B}\tau$.   In addition, a constant
stress along the $\hat{z}$ direction was applied during the cooling
process, while the normal stress components along the other two
directions were set to zero (see Fig.\,\ref{fig:snapshot_system}).
For reference, the simulations of the cooling process at zero
pressure along all three directions were also carried out. After
cooling, the applied stress was set to zero and the glass was
allowed to equilibrate during the additional time interval of
$10^4\,\tau$ at $T_{LJ}=0.01\,\varepsilon/k_B$ and $P=0$. Finally,
the glassy samples were strained at the constant rate of
$10^{-5}\,\tau^{-1}$ at $T_{LJ}=0.01\,\varepsilon/k_B$ and zero
pressure in order to evaluate the elastic modulus and the peak value
of the stress overshoot. The data for the potential energy and
mechanical properties were averaged over 15 independent samples.

\section{Results}
\label{sec:Results}


It is well understood by now that the potential energy of glasses
and, consequently, their mechanical properties depend sensitively on
the rate of cooling from the liquid state~\cite{Greer16}. In
general, more slowly cooled glass formers settle at lower energy
states, and, under external deformation, exhibit higher yield stress
and are more brittle~\cite{OHern17}.  Remarkably, it was recently
demonstrated experimentally that ductility can be enhanced when
sufficiently large normal stress is applied during the cooling
process, leading to highly rejuvenated states~\cite{Schroers20}. The
key parameters of the cooling under stress protocol include the
initial temperature of a glass former, the rate of cooling, and the
applied stress. In the present study, the binary mixture was cooled
under stress with the fixed rate of $10^{-4}\varepsilon/k_{B}\tau$.
We note that cooling at higher rates, $10^{-2}\varepsilon/k_{B}\tau$
and $10^{-3}\varepsilon/k_{B}\tau$, and zero pressure was shown to
produce highly rejuvenated glassy samples with already low values of
the yielding peak~\cite{Priez18tcyc,Priez19T5000}.  On the other
hand, test runs at the lower cooling rate,
$10^{-5}\varepsilon/k_{B}\tau$, did not reveal significant
rejuvenation, in part because only a relatively small stresses can
be applied during the cooling process without extended flow near the
glass transition temperature. Hence, for each initial temperature
near $T_g$, the values of the applied stress were chosen by trial
and error up to the maximum stress above which samples break.

%
\begin{figure}[t]
\includegraphics[width=12.0cm,angle=0]{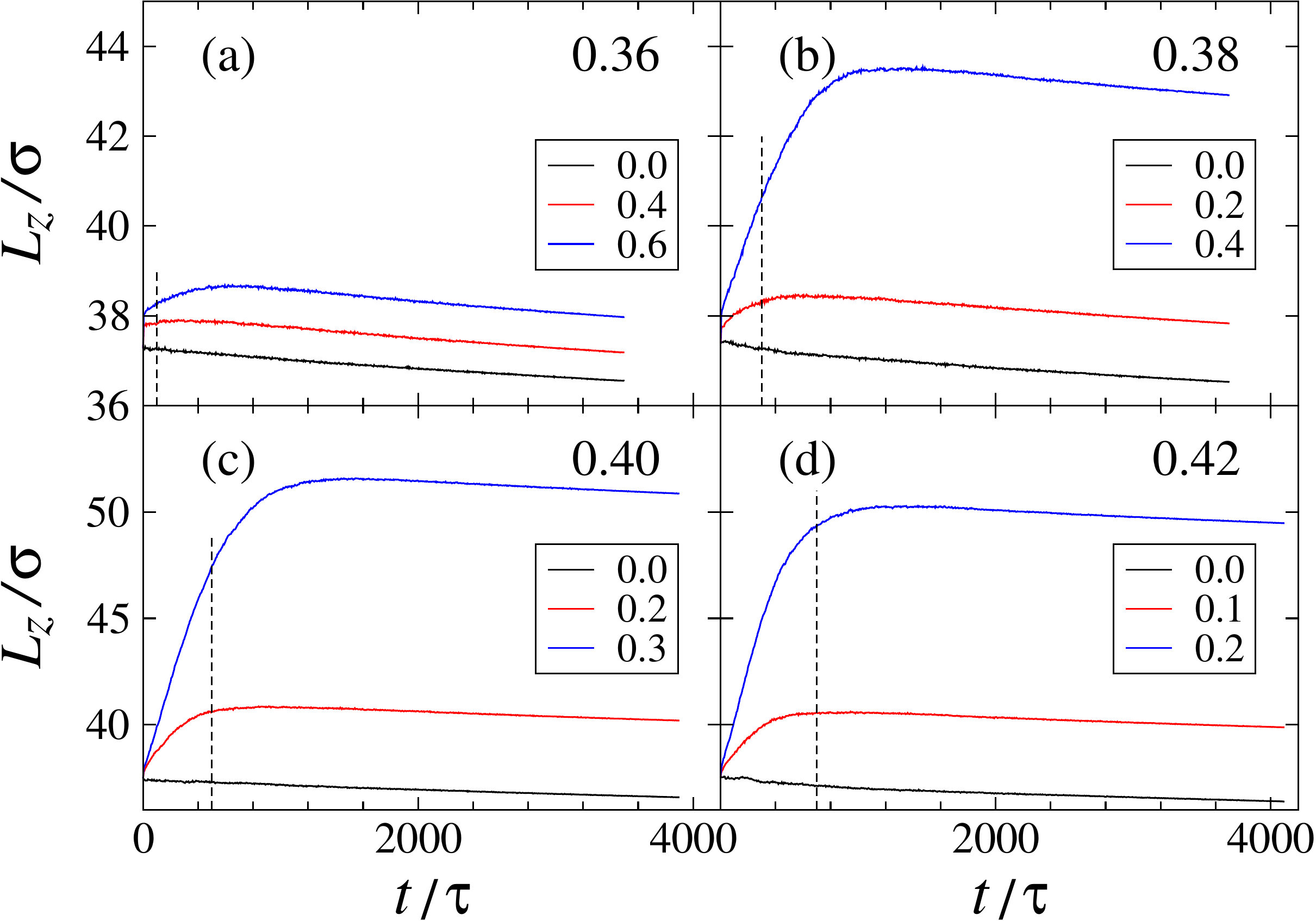}
\caption{(Color online) The system size along the $\hat{z}$
direction during cooling from the indicated initial temperatures to
$T_{LJ}=0.01\,\varepsilon/k_B$ with the rate
$10^{-4}\varepsilon/k_{B}\tau$. The values of the applied stress (in
units of $\varepsilon\sigma^{-3}$) are listed in the legends. The
vertical dashed lines denote times when the system temperature is
$0.35\,\varepsilon/k_B$. The scales along the vertical axes are
different on the upper and lower panels. }
\label{fig:Lz_4Ts}
\end{figure}

\vskip 0.05in


In our setup, the binary mixture is initially equilibrated near
$T_g$ in a cubic box with periodic boundaries, and then cooled under
stress applied along the $\hat{z}$ direction, which might result in
significant deformation of the computational domain. The time
dependence of the system size along the $\hat{z}$ direction is
reported in Fig.\,\ref{fig:Lz_4Ts} for the selected values of the
applied stress.  For each value of the initial temperature, the
variation of $L_z$ is shown for zero, intermediate value, and the
maximum applied stress above which samples undergo extensive plastic
deformation near the glass transition temperature.  It can be
clearly seen in Fig.\,\ref{fig:Lz_4Ts} that with increasing applied
stress, the deformation becomes more pronounced in the vicinity of
$T_g=0.35\,\varepsilon/k_B$, which is marked by the vertical dashed
lines.  Note that the negative slope at later times corresponds to
contraction upon cooling below the glass transition temperature,
where the applied stress becomes smaller than the yield stress.
Shown for reference in all panels in Fig.\,\ref{fig:Lz_4Ts}, the
black curves denote the variation of $L_z$ during the cooling
process at zero pressure for different initial temperatures. As
deduced from the blue curve in Fig.\,\ref{fig:Lz_4Ts}\,(c), the
maximum elongation upon cooling to $T_{LJ}=0.01\,\varepsilon/k_B$
under stress is $L_z/L_x\approx1.6$.

%
\begin{figure}[t]
\includegraphics[width=12.0cm,angle=0]{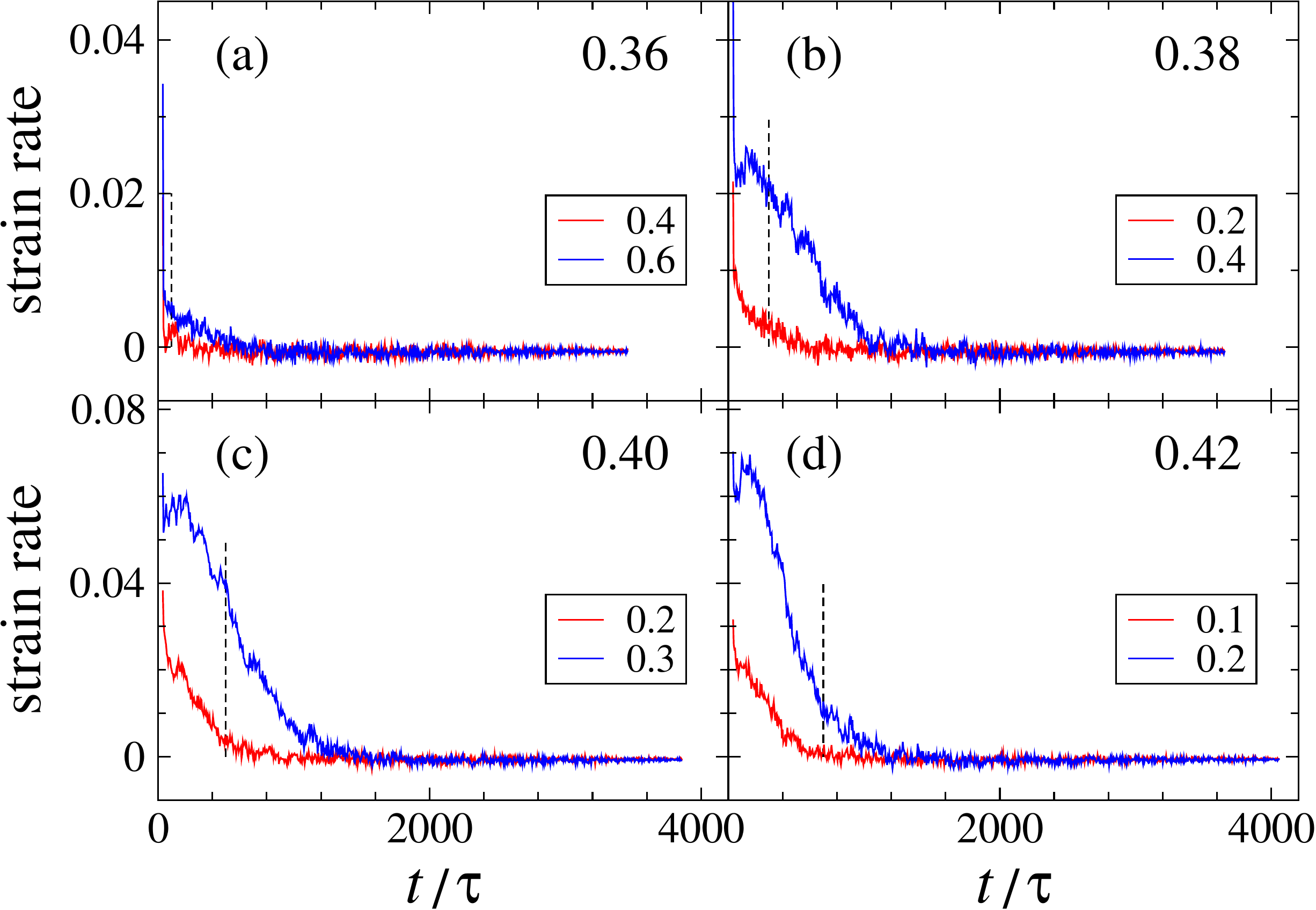}
\caption{(Color online) The rate of strain along the $\hat{z}$
direction (in units of $\tau^{-1}$) during cooling from the initial
temperatures to $T_{LJ}=0.01\,\varepsilon/k_B$ with the rate
$10^{-4}\varepsilon/k_{B}\tau$. The same data as in
Fig.\,\ref{fig:Lz_4Ts}. The initial temperatures are (a)
$0.36\,\varepsilon/k_B$, (b) $0.38\,\varepsilon/k_B$, (c)
$0.40\,\varepsilon/k_B$, and (d) $0.42\,\varepsilon/k_B$. The
applied stress (in units of $\varepsilon\sigma^{-3}$) along the
$\hat{z}$ direction is indicated in the legends. The vertical dashed
lines denote the temperature of $0.35\,\varepsilon/k_B$. Note that
the vertical scales are different on the upper and lower panels. }
\label{fig:strain_rate_4Ts}
\end{figure}

\vskip 0.05in


The rejuvenated structure in a strained sample is formed mostly
during the time interval when the system temperature is reduced in a
close vicinity of the glass transition temperature. Otherwise, a
liquid glass former flows above $T_g$, whereas in the glass phase,
the applied stress remains smaller than the yield stress, and the
effect of elastostatic loading on the energy state during
$\sim\!10^3\,\tau$ is negligible~\cite{PriezELAST19}. Similar to the
definition used in Ref.\,\cite{Schroers20}, the strain along the
$\hat{z}$ direction was computed as follows:
\begin{equation}
\varepsilon(t)=[L_z(t)-L_z(0)]/L_z(0)\cdot 100,
\label{Eq:strain}
\end{equation}
where $L_z(0)$ is the initial length of the computational domain and
$L_z(t)$ is the time dependent length.  The rate of strain as a
function of time is plotted in Fig.\,\ref{fig:strain_rate_4Ts} for
the same values of the applied stress and initial temperature as in
Fig.\,\ref{fig:Lz_4Ts}.  It can be readily observed that the rate of
strain at $T_g=0.35\,\varepsilon/k_B$ is larger at higher initial
temperatures and the maximum applied stress (marked by the blue
curves in Fig.\,\ref{fig:strain_rate_4Ts}). The exception to this
trend is the case of the initial temperature
$T_{LJ}=0.42\,\varepsilon/k_B$ where the maximum applied stress is
relatively low, $0.2\,\varepsilon\sigma^{-3}$, and most of the
deformation occurs just above $T_g$ [\,see
Fig.\,\ref{fig:strain_rate_4Ts}\,(d)\,]. Further, it was argued that
the effect of rejuvenation is enhanced when the typical timescale
associated with the cooling process (\textit{i.e.}, the inverse
cooling rate) exceeds the characteristic time of the straining
process, which is roughly a few percent divided by the strain
rate~\cite{Schroers20}. For the parameters of the present study,
this condition is satisfied for all values of the applied stress,
thus leading to rejuvenated states with respect to energy levels
obtained upon cooling at $P=0$.

%
\begin{figure}[t]
\includegraphics[width=12.0cm,angle=0]{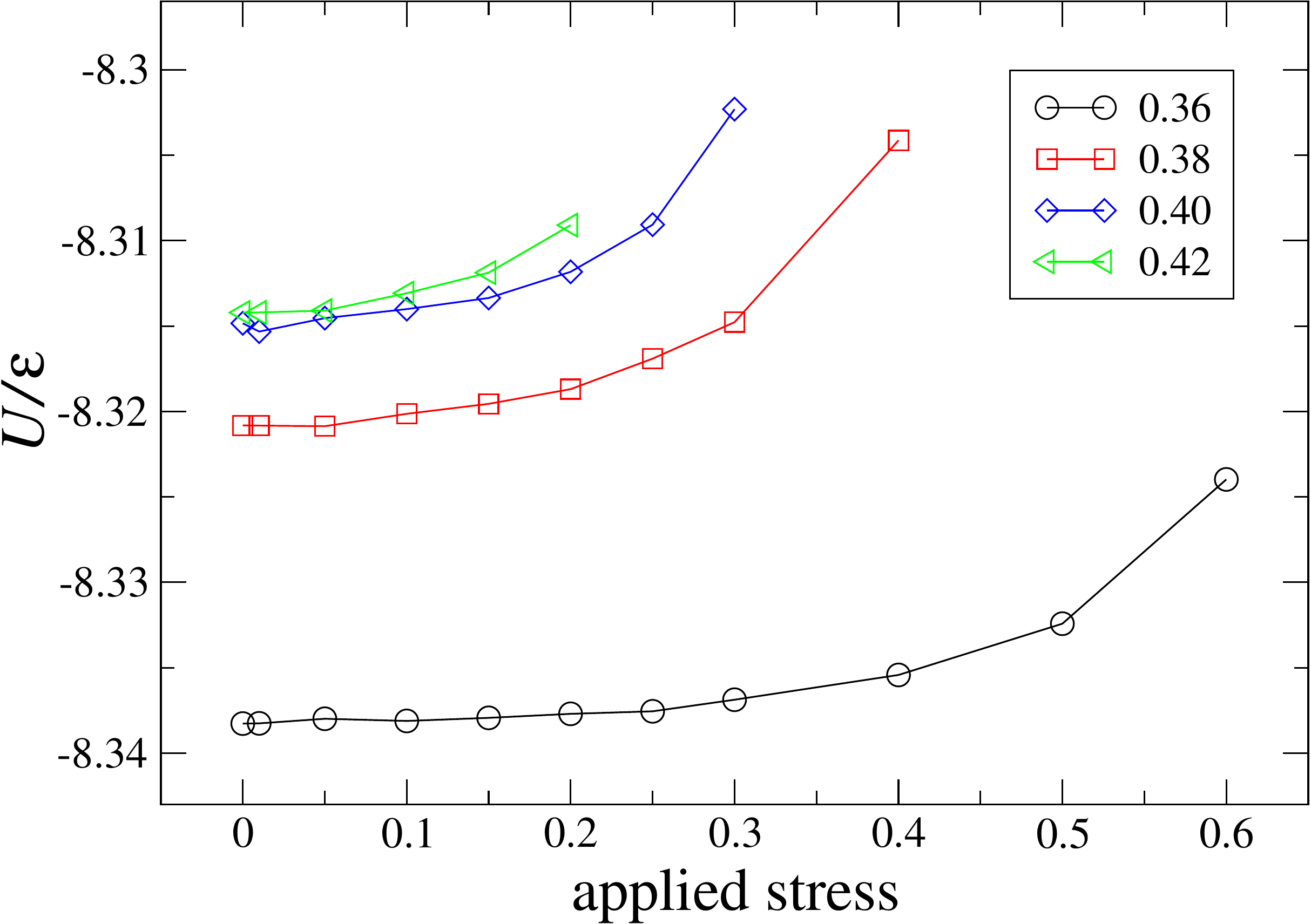}
\caption{(Color online) The potential energy per atom as a function
of the applied stress (in units of $\varepsilon\sigma^{-3}$) during
cooling to the temperature $T_{LJ}=0.01\,\varepsilon/k_B$ with the
rate $10^{-4}\varepsilon/k_{B}\tau$. The values of the initial
temperature (in units of $\varepsilon/k_B$) are listed in the
legend. The potential energy is computed at
$T_{LJ}=0.01\,\varepsilon/k_B$ and zero pressure. The data were
averaged over 15 independent samples. }
\label{fig:poten_4Ts_Pzz}
\end{figure}

\vskip 0.05in


Next, the variation of the potential energy per atom as a function
of stress applied during the cooling process is shown in
Fig.\,\ref{fig:poten_4Ts_Pzz} for the initial temperatures
$T_{LJ}=0.36\,\varepsilon/k_B$, $0.38\,\varepsilon/k_B$,
$0.40\,\varepsilon/k_B$, and $0.42\,\varepsilon/k_B$.  The data in
Fig.\,\ref{fig:poten_4Ts_Pzz} were averaged over 15 samples after
annealing during $10^4\,\tau$ at $T_{LJ}=0.01\,\varepsilon/k_B$ and
$P=0$.   It can be seen that for zero applied stress, the potential
energy is reduced upon decreasing initial temperature.  It was
previously demonstrated that the potential energy of the inherent
structures of the KA mixture at zero pressure is reduced when the
annealing temperature approaches $T_g$ from
above~\cite{Priez20tfic}. Further, with increasing applied stress,
the binary glass is relocated to progressively higher energy states
as it freezes at higher strain. It should be commented that the
maximum increase in the potential energy at
$T_{LJ}=0.36\,\varepsilon/k_B$ and $0.38\,\varepsilon/k_B$ is
slightly larger than the most pronounced rejuvenation due to
elastostatic loading of the KA glass detected at about half the
glass transition temperature~\cite{PriezELAST19,PriezELAST21}. On
the other hand, it was shown for the same binary mixture model that
rejuvenation is significantly enhanced as a result of flush
annealing when samples are rapidly heated above $T_g$ and cooled at
effectively high rates into the glass phase~\cite{Priez19one}.

%
\begin{figure}[t]
\includegraphics[width=12.0cm,angle=0]{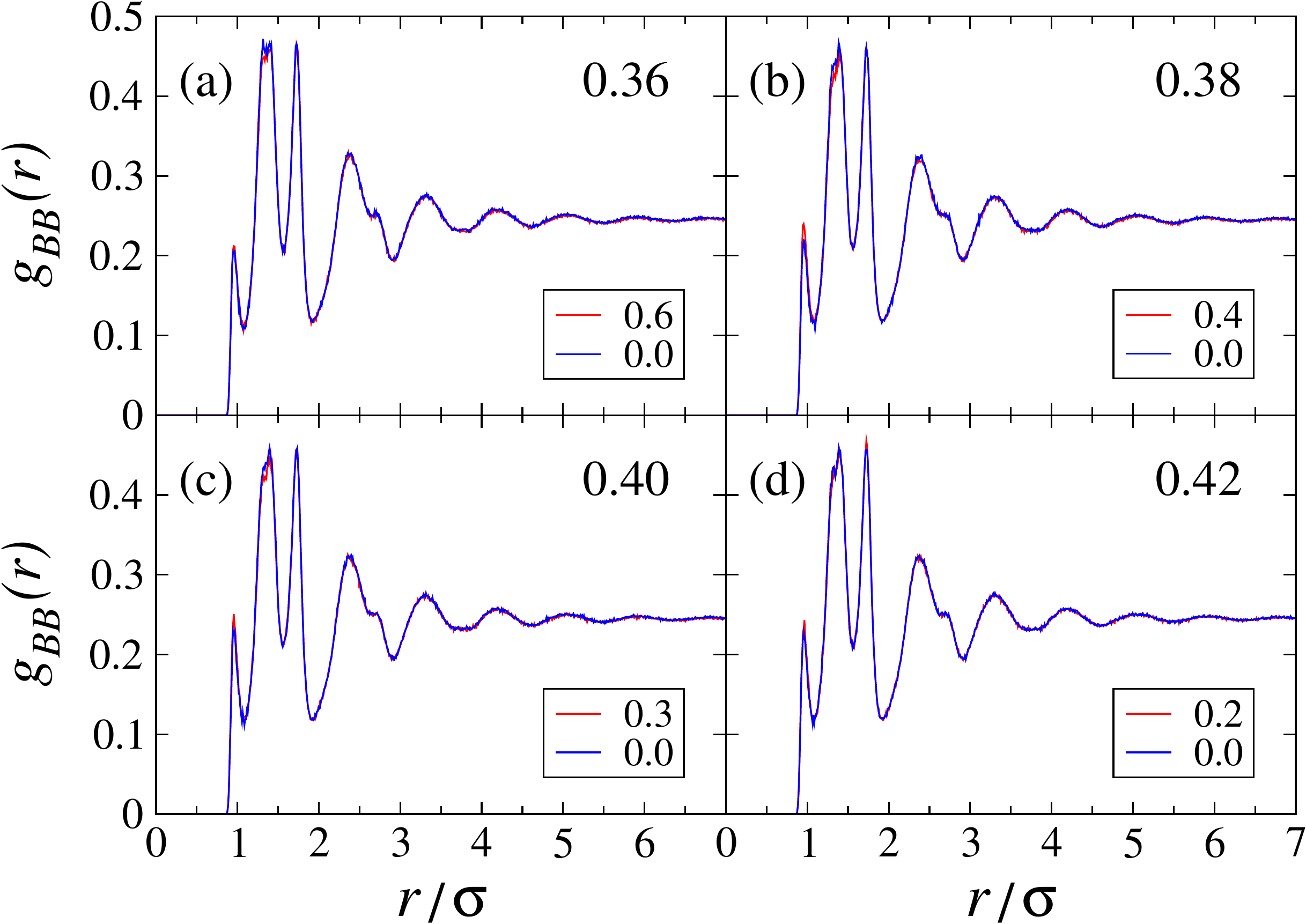}
\caption{(Color online) The radial distribution function,
$g_{BB}(r)$, after cooling to $T_{LJ}=0.01\,\varepsilon/k_B$ from
the initial temperatures (a) $0.36\,\varepsilon/k_B$, (b)
$0.38\,\varepsilon/k_B$, (c) $0.40\,\varepsilon/k_B$, and (d)
$0.42\,\varepsilon/k_B$. The values of the applied stress (in units
of $\varepsilon\sigma^{-3}$) along the $\hat{z}$ direction are
listed in the legends. }
\label{fig:grBB}
\end{figure}

\vskip 0.05in


The increase in potential energy reported in
Fig.\,\ref{fig:poten_4Ts_Pzz} is reflected in the atomic structure,
which can be analyzed using the radial distribution function.  Thus,
it was previously found that the most sensitive measure of the
structural changes in the KA model glass is the average separation
between neighboring atoms of type
$B$~\cite{Vollmayr96,Stillinger00,Priez20tfic}.    This correlation
can be understood by realizing that the interaction energy between
smaller atoms of type $B$ in the LJ potential,
Eq.\,(\ref{Eq:LJ_KA}), is the lowest among $\varepsilon_{AA}$,
$\varepsilon_{AB}$, and $\varepsilon_{BB}$, leading to a reduced
number of contacts between $B$ type atoms in a well annealed (low
energy) glass~\cite{Vollmayr96,Stillinger00}. The averaged radial
distribution function, $g(r)_{BB}$, is shown in Fig.\,\ref{fig:grBB}
for zero and maximum applied stress during cooling from the
indicated initial temperatures to $T_{LJ}=0.01\,\varepsilon/k_B$. It
can be observed that upon increasing stress, the height of the first
peak at about $0.96\,\sigma$ becomes slightly larger and the
magnitude of the second peak at $\approx\!1.38\,\sigma$ is reduced.
This trend is consistent with the increase in potential energy when
samples are cooled under stress, thus leading to a more random
packing and larger number of contacts between $B-B$ type atoms.

%
\begin{figure}[t]
\includegraphics[width=12.0cm,angle=0]{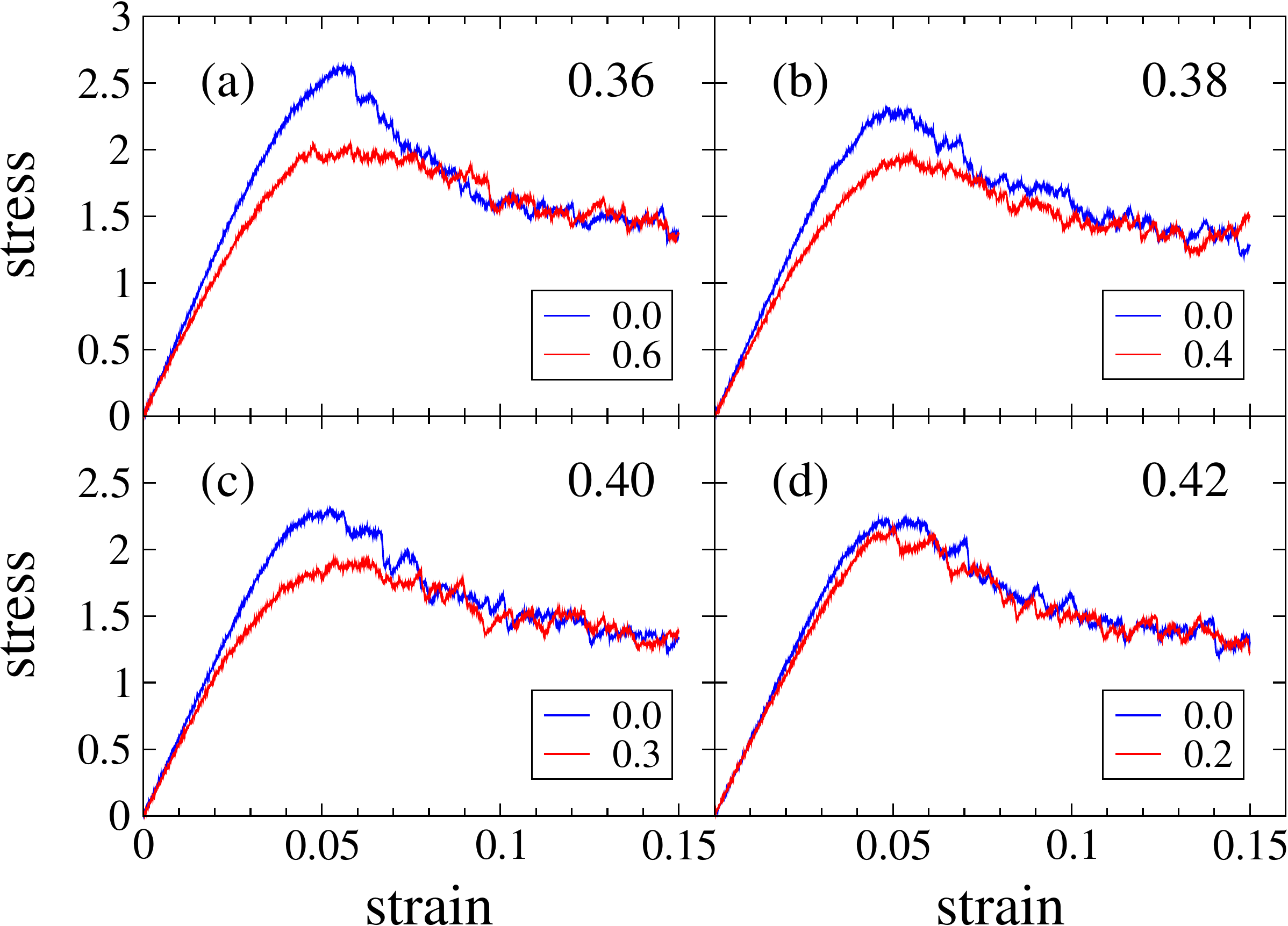}
\caption{(Color online) The variation of the normal stress (in units
of $\varepsilon\sigma^{-3}$) as a function of strain along the
$\hat{z}$ direction during steady loading with the rate
$10^{-5}\,\tau^{-1}$ at $T_{LJ}=0.01\,\varepsilon/k_B$ and zero
pressure.  The samples were prepared by cooling from the initial
temperatures (a) $0.36\,\varepsilon/k_B$, (b)
$0.38\,\varepsilon/k_B$, (c) $0.40\,\varepsilon/k_B$, and (d)
$0.42\,\varepsilon/k_B$.  The values of the applied stress during
the cooling process are listed in the insets.  See text for
details.}
\label{fig:stress_strain}
\end{figure}

\vskip 0.05in


In order to evaluate changes in mechanical properties due to cooling
under applied stress, the glassy samples were strained at the
constant rate of $10^{-5}\,\tau^{-1}$ at
$T_{LJ}=0.01\,\varepsilon/k_B$ and zero pressure.  The
representative stress-strain curves are plotted in
Fig.\,\ref{fig:stress_strain} for the indicated values of the
initial temperature.   In each case, the data are reported for zero
and maximum applied stress during the cooling process. As is
evident, the height of the yielding peak increases in samples
prepared at lower initial temperatures and zero external stress.
This is consistent with the results in Fig.\,\ref{fig:poten_4Ts_Pzz}
where more relaxed states were obtained by cooling at zero applied
stress and lower temperatures.  Moreover, it is clearly seen that
the yielding peak in Fig.\,\ref{fig:stress_strain} is reduced in
highly rejuvenated samples initially cooled at the maximum applied
stress. Hence, it can be concluded that the maximum difference in
the yield stress becomes more pronounced for binary glasses prepared
at lower initial temperatures.

%
\begin{figure}[t]
\includegraphics[width=12.cm,angle=0]{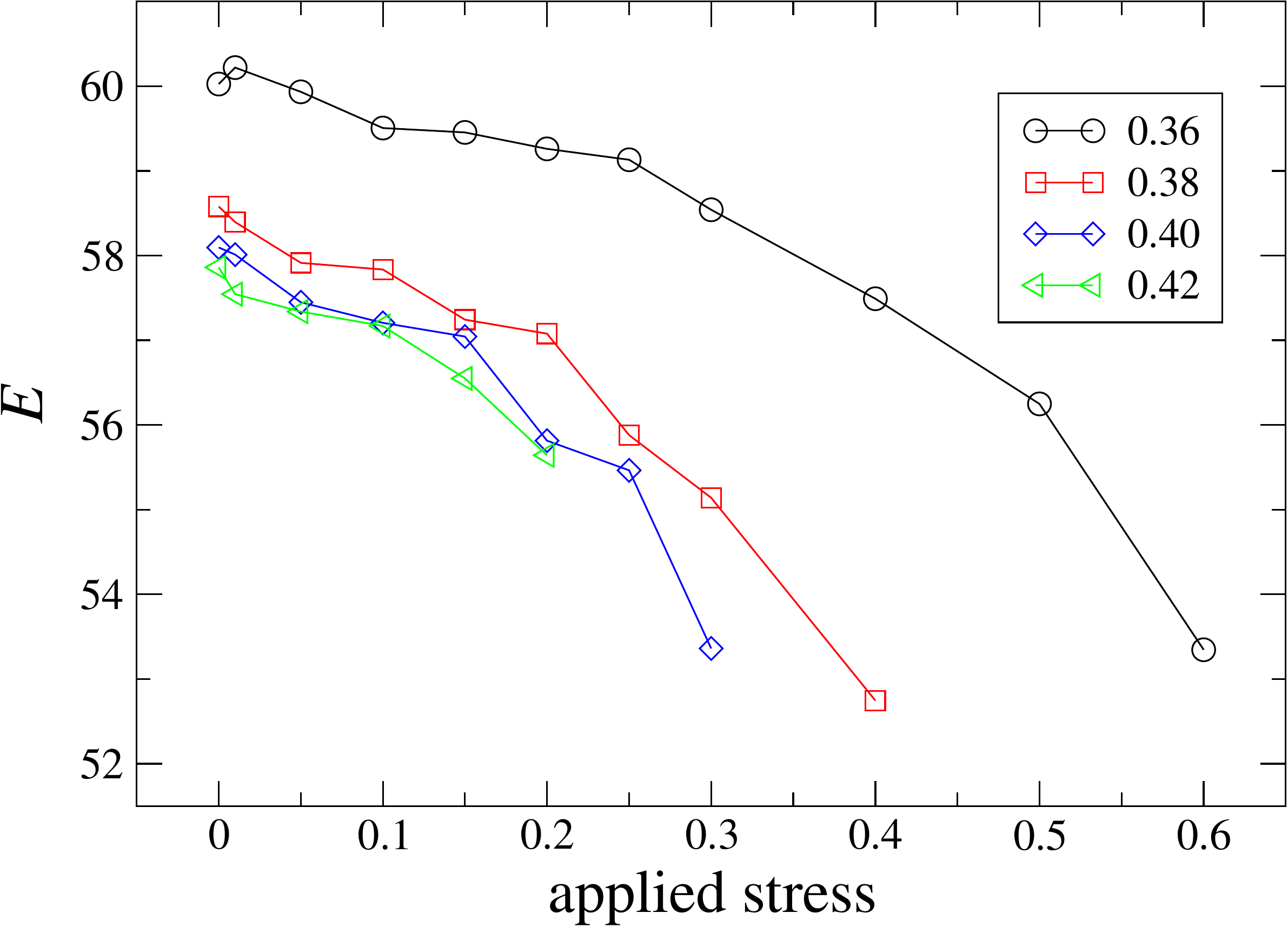}
\caption{(Color online) The elastic modulus $E$ (in units of
$\varepsilon\sigma^{-3}$) versus the applied stress (in units of
$\varepsilon\sigma^{-3}$) used during the cooling process for the
tabulated values of the initial temperature. The elastic modulus was
measured during tension along the $\hat{z}$ direction with the rate
$10^{-5}\,\tau^{-1}$ at $T_{LJ}=0.01\,\varepsilon/k_B$ and zero
pressure. }
\label{fig:E}
\end{figure}

\vskip 0.05in

\begin{figure}[t]
\includegraphics[width=12.cm,angle=0]{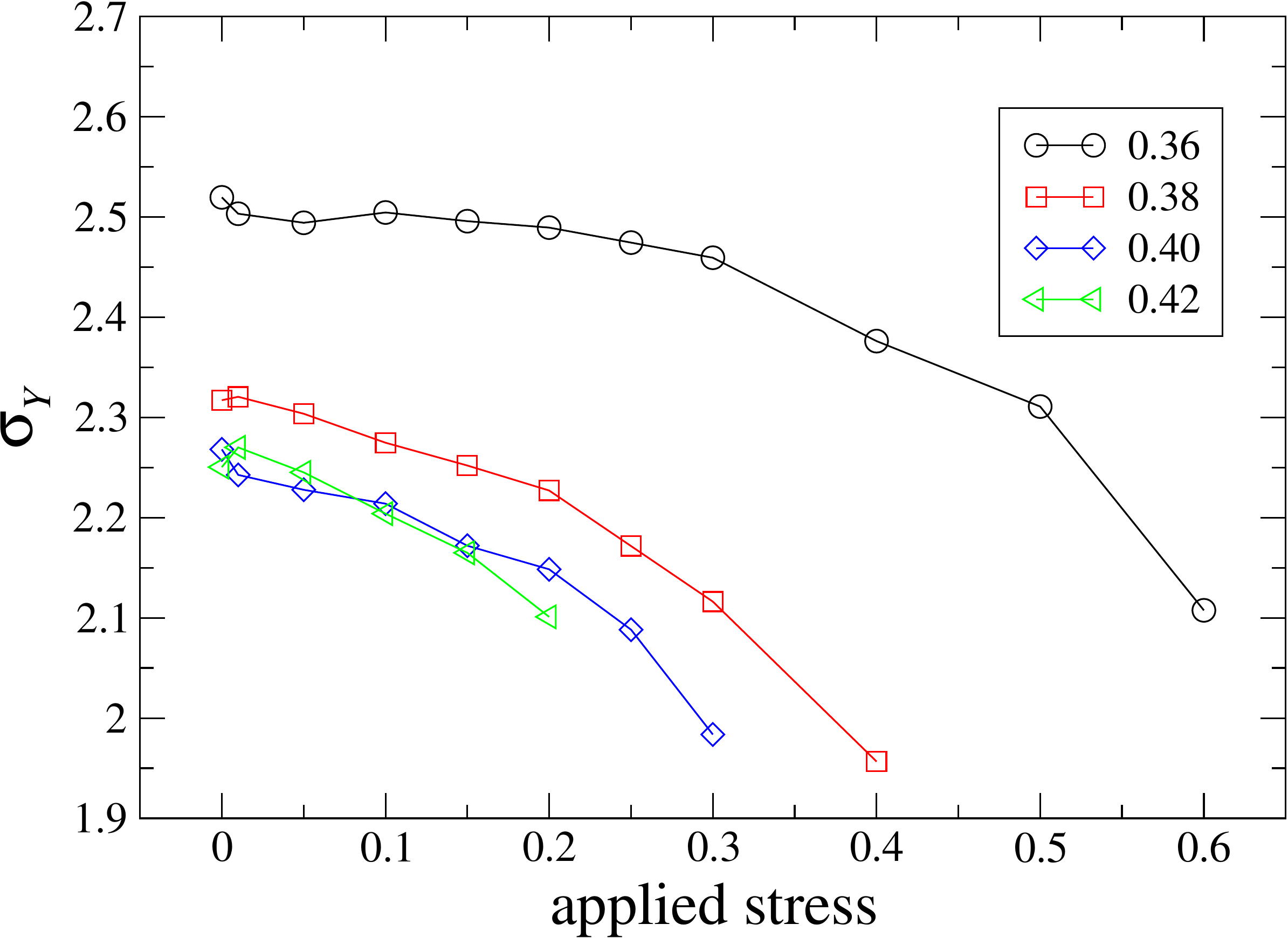}
\caption{(Color online) The peak value of the stress overshoot
$\sigma_Y$ (in units of $\varepsilon\sigma^{-3}$) as a function the
applied stress (in units of $\varepsilon\sigma^{-3}$) used during
cooling to $T_{LJ}=0.01\,\varepsilon/k_B$ for the indicated values
of the initial temperature. The same samples and loading conditions
as in Fig.\,\ref{fig:E}. }
\label{fig:sigY}
\end{figure}

The simulation results for the elastic modulus, $E$, and the peak
value of the stress overshoot, $\sigma_Y$, in steadily strained
samples at $T_{LJ}=0.01\,\varepsilon/k_B$ are presented in
Figures\,\ref{fig:E} and \ref{fig:sigY}, respectively. Here, the
data were averaged over 15 realizations of disorder. For each
sample, the elastic modulus was computed from the best linear fit to
stress at small values of strain, $\varepsilon_{zz}\leqslant0.01$,
and the yielding peak represents the maximum of the stress-strain
curves in the range $\varepsilon_{zz}\leqslant0.15$. As expected,
both $E$ and $\sigma_Y$ tend to increase in better annealed samples
that were cooled from lower initial temperatures at zero stress (see
Figs.\,\ref{fig:E} and \ref{fig:sigY}). Except for the case
$T_{LJ}=0.42\,\varepsilon/k_B$, the elastic modulus is reduced by
about 10\% and the yielding peak is decreased by roughly 15\%, when
the applied stress varies from zero to a maximum value. All in all,
these results demonstrate that the process of cooling from initial
temperatures near the glass transition temperature and concomitantly
applying a constant stress in tension leads to improved plasticity
in amorphous alloys.

%

\section{Conclusions}

In summary, we studied the effect of strain rate applied during
cooling of amorphous alloys on their energy states and mechanical
properties using atomistic simulations. The model glass was
represented via a binary mixture which was first equilibrated near
the glass transition temperature and then linearly cooled under
constant tensile stress deep into the glass phase. It was found that
the potential energy of binary glasses increases when a larger
tensile stress was applied during the cooling process, and, as a
result, higher strain rate during freezing near the glass transition
temperature.   In turn, the maximum value of the applied stress is
reduced at higher initial temperatures since glassy samples can be
extensively deformed above $T_g$.  Furthermore, it was shown that
the amorphous structure of rejuvenated glasses contains a larger
number of contacts between smaller atoms, as reflected in the shape
of the radial distribution function. Lastly, the ductility is
enhanced as the elastic modulus and the yielding peak are reduced in
glasses that were cooled at larger tensile stresses and higher
initial temperatures.

\section*{Acknowledgments}

Financial support from the National Science Foundation (CNS-1531923)
and the ACS Petroleum Research Fund (60092-ND9) is gratefully
acknowledged. The author would like to thank Prof. J. Schroers for
useful comments. The article was prepared within the framework of
the HSE University Basic Research Program and funded in part by the
Russian Academic Excellence Project `5-100'. The molecular dynamics
simulations were performed using the LAMMPS open-source code
developed at Sandia National Laboratories~\cite{Lammps}. The
numerical simulations were carried out at the Wright State
University's Computing Facility and the Ohio Supercomputer Center.



\bibliographystyle{prsty}

\end{document}